\documentclass{elsart}
\usepackage{graphicx}
\usepackage{amssymb}
\usepackage[fleqn]{amsmath}
\begin{document}

\begin{frontmatter}
\title{A novel scheme for entanglement engineering in a fermionic system}
\author{Jiming Gao,}
\author{Jiaxiang Wang\corauthref{cor1}}
\corauth[cor1]{corresponding author}
\address{Institute of Theoretical Physics, Department of Physics, East China Normal
University, Shanghai 200062}
\address{China State Key Laboratory of
Precision Spectroscopy (East China Normal University)}
\ead{jxwang@phy.ecnu.edu.cn}
\begin{abstract}
In this letter, we present a novel scheme to engineer the
entanglement in a fermionic system, which is modeled by a minimally
three-site Hubbard model. It is found that, in this type of system,
we can have two free parameters. One is used to tune the
entanglement and the other to switch on or off the above tuning
function. The whole process is much like what happens in a
traditional transistor switch, where the main current between the
emitter and collector can be turned on or off by the base current.
\end{abstract}

\end{frontmatter}

Quantum entanglement, is considered to be one of the key concepts in
quantum mechanics, which has no classical analog. Ever since first
noted by Einstein, Podolsky, and Rosen (EPR) \cite{Einstein} and
Schr\"odinger \cite{Schr}, the entanglement has received wide
experimental and theoretical attentions
\cite{bell,Bennett,Hagley,Turchette,Bouwmeester,Monroe}. Recently,
it is believed that quantum entanglement is a crucial resource in
quantum information processing (QIP) \cite{Nielsen}, such as quantum
computations, quantum state transfer \cite{Li}, dense coding
\cite{Bennetts,Mattle}, quantum communication \cite{Schumacher} and
quantum cryptography \cite{Bennettg}. Hence the present research
upon entanglement has moved from philosophical debates to applied
fields and concrete theoretical study. And generally speaking, the
entanglement "engineering" has become the core of QIP, no matter
whether it is realized with optical systems or with atomic ones.

In condensed matter physics, correspondingly, much of the
theoretical investigations are focused upon spin or itilerant
fermionic systems, such as the transverse Ising model
\cite{Osterloh}, Heisenberg model \cite{benjamin} and Hubbard model
\cite{Zanardi}. A marvelous progress in this direction is the
discovery of the close relationship between the entanglement and the
quantum phase transition (QPT) \cite{Osterloh,gu,wang}. It has been
found that there is an abrupt change of the entanglement over the
quantum critical point. The significance of this result lies not
only in its physical implication to describe the long-range
correlations in the many-body systems, but also in its potential
application to control the entanglement in QIP. For example, we can
use this effect to realize an entanglement switch by moving the
system from one side of the critical point to the other side. During
this process, the entanglement will jump between two stable values,
just like a switch. This nice idea has been put forward for some
time now \cite{Osterloh,wang,wangJ,Osenda}. But in its practicle
application, we can observe two obstacles. Firstly, QPT only happens
in thermodynamic limit, which means that a large number of particles
will be involved in the process. This will bring much trouble in
pinning down the entanglement among finite number of particles or
sites. Secondly, the infinite size of the system will put a sever
limit upon the response time of the entanglement switch, which will
influence the whole QIP efficiency. These two problems are inherent
in QPT and hard to be solved so long as QPT is involved. But the
idea of entanglement switch has motivated us to put forward a new
scheme, which could not only realize the same purpose as QPT has
been intended to achieve, but also give us more. In the following, a
tight-binding Hubbard model will be used as a generic example to
illustrate our idea.

For QPT, the principal trouble for efficient engineering of the
entanglement lies in its infinite number of freedoms. Hence, in our
model, we will only consider finite number of sites. Moreover if
only the engineering of the well-understood two-particle
entanglement is concerned, we need to consider at least a two-site
Hubbard model. In the past few years, torrents of experimental and
theoretical work have been carried out for two-site systems under
various conditions \cite{Zhang,Petta,ZhangP,Rycerz}. But for our
purpose, as we shall see, two sites are not enough to realize the
switch function. Hence, an extra site will be included, which plays
a crucial role in controlling the entanglement engineering between
the other two sites. Actually, this kind of three-site system can
demonstrate fruitful static and dynamic effect as shown in Ref.
\cite{ivanchenko} for a spin chain. Here, for the 3-site Hubbard
model, we will not investigate all its quantum behavior, but focus
upon the effect of entanglement transistor switch as we name it.
Before going to the details, the basic engineering process will be
summarized as follows. Namely, two free parameters will be chosen
first. Then one is used to switch or tune the entanglement and the
other to turn off or on the above function. We call these two
parameters as "tuning parameter" and "control parameter",
respectively. Compared with QPT, this is a few-body system, in which
the two-site entanglement could be well defined and analyzed for
QIP. Most importantly, the response time will be much reduced.

Fig. 1 schematically shows the system setup in two different
geometries. For the cyclic geometry in Fig. 1(a), the The
Hamiltonian is,
\begin{equation}
H =  - t\sum\limits_\sigma  {\left( {c_{1\sigma }^\dag  c_{2\sigma }
+ c_{2\sigma }^\dag  c_{3\sigma } +c_{3\sigma }^\dag  c_{1\sigma } +
h.c.} \right) + U_0 \left( {n_{1 \uparrow } n_{1 \downarrow }  +
n_{2 \uparrow } n_{2 \downarrow } } \right) + Un_{3 \uparrow } n_{3
\downarrow } },\label{hamiltonian1}
\end{equation}
where $t$ stands for the hopping between the nearest neighboring
sites, $\sigma=\uparrow,\downarrow$ is the electron spin,
$c^{\dagger}_{i\sigma}$ and $c_{j\sigma}$ are the creation and
annihilation operators on the $i$th site and $U$ and $U_{0}$ are the
on-site Coulomb interactions. For the linear chain geometry in Fig.
1(b), the Hamiltonian is a bit different,
\begin{equation}
H =  - t\sum\limits_\sigma  {\left( {c_{1\sigma }^\dag  c_{2\sigma }
+ c_{2\sigma }^\dag  c_{3\sigma }  + h.c.} \right) + U_0 \left(
{n_{1 \uparrow } n_{1 \downarrow }  + n_{2 \uparrow } n_{2
\downarrow } } \right) + Un_{3 \uparrow } n_{3 \downarrow }
},\label{hamiltonian2}
\end{equation}
In fact, the two geometries are equivalent to two different boundary
conditions. Fig. 1(a) is for a periodic boundary condition and Fig.
1(b) for a open boundary one. The entanglement between the first two
sites inside the dotted box is what we intend to engineer. And the
parameters related to the first two sites are assumed to be the
same. Due to the itilerant feature of the system, each site has four
possible states, which could be taken as a generalized version of
two-state qubit. The four states can be expressed as $\mid
0{\rangle},\mid \uparrow{\rangle},\mid \downarrow{\rangle},\mid
\uparrow\downarrow{\rangle}$. Depending upon the number and the
spins of the electrons contained in the system, some of the four
states might be depressed. By convention, for this kind of itinerant
fermionic system, the von Neuman entropy will be used to measure the
entanglement, i.e. if a two-part system is described by a density
matrix $\rho$, the entanglement between the two parts can be
measured by the following expression,
\begin{equation}
E=-Tr(\rho_{1}log_{2}\rho_{1}), \label{definition}
\end{equation}
where $\rho_{1}=Tr_{2}(\rho)$ denotes the tracing over the freedoms
from the second part of the system.

From Eq. (\ref{hamiltonian1}) and Eq. (\ref{hamiltonian2}), it is
also apparent that, if we scale all the energies by $t$, only two
free parameters are left, i.e. $U/t$ and $U_0/t$. To see the role
played by site 3, which provides us the control function through the
parameter $U/t$, we first cut it off from the system. In this
situation, the density matrix $\rho$ can written as $ \left| {\psi
_0 } \right\rangle \left\langle {\psi _0 } \right|$ with $\psi _0$
to be the ground state wave function. In order to have a system as
simple as possible, we also need to fix the particle number $N$ and
the total electron spin $S_z$, both of which are good quantum
numbers. The cases with $N=0$, $N=1$, $N=3$ and $N=4$ are trivial
since the corresponding ground state is independent of $U_0/t$,
which is also true for $N=2$ with $S_z=\pm1$. Hence the only
nontrivial case is for $N=2$ with $S_z=0$, i.e., one electron spin
up and the other down. For this well-defined case, the numerical
method of exact diagonalization is used to calculate the
entanglement. The results are presented in Fig.2(a). It can be
easily seen that $U_0/t=0$ corresponds to the maximal entanglement
with $E=2$, which is a direct consequence of equal superposition of
the configuration $\left| {\left( { \uparrow \downarrow } \right)_1
(0)_2 } \right\rangle ,\left| {(0)_1 \left( { \uparrow  \downarrow }
\right)_2 } \right\rangle ,\left| { (\uparrow) _1  (\downarrow) _2 }
\right\rangle $ and $\left| { (\downarrow) _1 (\uparrow) _2 }
\right\rangle$ in the ground state. Here $($ $)_i$ denotes the basis
state of the $i$th site.  The corresponding single-site density
matrix is
\begin{equation}
\centering \rho _1  = \frac{1}{4}\left( {\begin{array}{*{20}c}
   1 & 0 & 0 & 0 \\
   0 & 1 & 0 & 0 \\
   0 & 0 & 1 & 0 \\
   0 & 0 & 0 & 1 \\
\end{array}} \right)
\end{equation}
As $U_0/t\rightarrow\infty$, the double occupancy is forbidden due
to the strong on-site repulsion and the ground state is a
superposition of $ \left|{ (\uparrow) _1 (\downarrow) _2 }
\right\rangle $ and $ \left| { (\downarrow) _1 (\uparrow) _2 }
\right\rangle $ with equal coefficient, which leads to $E=1$.
Similarly, as $U_0/t\rightarrow-\infty$, the single occupancy is
depressed and only  $ \left|{(0)_1(\uparrow\downarrow)_2 }
\right\rangle $ and  $ \left|{ (\uparrow\downarrow)_1 (0)_2 }
\right\rangle $ are equally superposed, which also leads to $E=1$.
All these results are consistent with those in Ref.(\cite{Zanardi}).
It should be noted that, although in this case the entanglement can
be continuously tuned from minimum $1$ to maximum $2$,no switch
function as found in QPT can be realized since the entanglement
demonstrates no distinct saturation value.

The whole scenario changes when a third site is included. The
electron number $N$ and the total spin $S_z$ keeps the same as
above, namely, we still have two electrons with one electron spin up
and one spin down. Now the left two-site subsystem will be actually
in a mixed state if we reduce site 3 from the density matrix
$\rho=Tr_3 \left| {\phi_0 } \right\rangle \left\langle {\phi _0 }
\right|$, where $\left| {\phi_0 } \right\rangle$ is the ground state
of the whole 3-site system. Fig. 2(b) shows a typical result for
positive control parameter $U/t$. A great difference from Fig. 2(a)
is that the entanglement now displays two distinct saturation values
for positive $U/t$, just like what happens in QPT. This is exactly
what we are looking for, from which an entanglement switch can be
realized. And within a narrow region around $U_0/t=0$, $E$ can be
continuously tuned. The saturation values for the cyclic case can be
explained as follows. First, when $U_0/t$ is positive enough,
satisfying $U_0/t>U/t$, all the double occupancy on the three sites
are disfavored in energy. Then the two electrons will sit on
different sites with all possibilities. Due to the spatial symmetry
of the system, each site has equal probabilities to be in the states
$ \left| \uparrow\right\rangle $,$ \left| \downarrow\right\rangle
$,$ \left| 0\right\rangle $. Hence, the corresponding entanglement
is $E=log_23=1.58$. By similar reasoning, we can have $E=1$ when
$U_0/t\ll U/t$ due to the equal probability for each site to b in
the states of $ \left| \uparrow\downarrow\right\rangle $ and $
\left| 0 \right\rangle$. In the linear chain case, the same feature
exists except that the saturation values are a bit different, which
results from the asymmetric properties of the three sites.

Besides the above discussed switch function, another interesting
static characteristic is shown in Fig. 2(c). Namely, as $U/t$ is
negative, $E$ keeps to be zero in a wide region of $U_0/t$, the
width of which depends upon the control parameter $U/t$. Generally,
$U_0/t>U/t$ is required to make this phenomenon appear. This is a
consequence of the competition among the on-site interactions at
different sites. Namely, if the on-site attractions at site 3
overwhelms the attraction at site 1 and 2, the two electrons will
tend to be localized on site 3. So for the left two sites, there is
only one possible state, i.e. $\left |(0)_1(0)_2\right\rangle$,
which naturally leads to zero entanglement.

Through the above analysis, it can be seen that the role played by
site 3 is crucial. Depending upon the sign and magnitude of the
control parameter $U/t$, it can either act as a barrier forbidden
the double occupancy or acts as an attractor freezing both the
electrons on itself, thus providing different boundary conditions
for the left two sites. By utilizing this kind of constraints, a
novel scheme to engineer the entanglement between the left two sites
can be formed. First, when $U/t$ is positive enough, for example
$U/t\sim80$, the entanglement can be tuned between two different
saturation values by varying $U_0/t$. Then by letting $U/t$ negative
enough, for example $U/t\sim-80$, the system will be driven into a
cut-off status with zero entanglement. Because the whole picture is
much like what happens in a traditional transistor switch in
electronics, we called it "transistor switch" effect of the
entanglement. Schematically, the on-site interaction $U/t$ on site 3
can be compared to the base current and the entanglement to the main
current between the emitter and the collector. Hence, the tuning
function of the entanglement through $U_0$ can be switched on or off
by setting $U/t$ at different values, just as the flow of the main
current can be switched on or off by properly setting the base
current. This control mechanism can find potential applications in
QIP. For example, it can be used as a memory or storage media of
quantum information characterized by the entanglement. One of the
advantages of this scheme is that the different properties of the
entanglement make the whole mechanism very robust against the
fluctuations of the tuning parameter $U_0/t$. Moreover, the stored
quantum information can be easily erased just by letting $U/t$ at a
value more negative than $U_0/t$.

So far, we have only investigated a limited region of the parameter
space of $U_0/t$ and $U/t$ for our purpose. The entanglement
variation in the whole parameter space is plotted in Fig. 3(a) for
cyclic case. The results for the linear-chain case are similar and
not shown. From the figure, we can easily see how the transistor
switch effect comes out when the parameters are independently varied
along a specific route on the $U/t-U_0/t$ plane. Actually, if we can
change $U_0/t$ and $U/t$ together, different forms of the
entanglement variations can be realized, which will lead to various
interesting engineering mechanism of the entanglement.

As we know, for any quantum device, the decoherence problem is a
must to be discussed. Since we are using a toy model to illustrate
the main idea, we will only discussed the temperature-induced
decoherence here. Once the thermal effect is considered, the measure
of the entanglement becomes,
\begin{equation}
E = Tr_{2,3} \left[ {\frac{{\sum\limits_i {e^{ - \frac{{\varepsilon
_i }} {{kT}}} \left| {\varphi _i } \right\rangle \left\langle
{\varphi _i } \right|} }} {{\sum\limits_i {e^{ - \frac{{\varepsilon
_i }} {{kT}}} } }}} \right],
\end{equation}
where $T$ is the temperature, $k$ is the Boltzman constant,
$\left|\varphi_i \right \rangle$ denotes the eigenfunction of the
whole system with $\epsilon_i$ to be the corresponding eigenvalues
and $Tr_{2,3}$ means the reduction of the freedoms related to site 2
and site 3. It is apparent that as $T\rightarrow0$, only the ground
state contributes to the sum and this definition goes to Eq.
(\ref{definition}), as expected. The results are given in Fig. 3(b)
and 3(c) for two different temperatures. It can be seen that the
entanglement does not change much when the temperature is low
enough, i.e. $kT/t\ll U/t$ or $ U_0/t $. This is quite
understandable since we are talking about a ground state property
and the finite energy gap to the first excited state will help to
guarantee some robustness of the entanglement over the thermal
fluctuations. As we increase the temperature from zero, the biggest
modification always starts from the border between the saturation
plateaus and then extend to the other regions. This is due to the
fact that the saturation border normally runs along $U/t=0$ or
$U_0/t=0$, which sets the energy upper threshold for the ground
state stability to a relatively low scale. And as
$T\rightarrow\infty$, all the possibilities of the states will be
excited by the thermal fluctuations. Hence the entanglement goes to
a flat maximum, making all the engineerings impossible.

As mentioned before, what we discussed here is a toy model. But due
to the rapid technology progress related to quantum dots
\cite{Recher} and especially cold atoms in optical lattices
\cite{anderlini}, people have been able to readily manipulate single
cold atoms in a well-controlled optical lattice or single ion and
electron in different kinds of wells. For example, in
Ref.(\cite{georges}), the system with cold atoms loaded into a
optical lattice is used to mimic Hubbard model. The experimentally
controllable parameters, such as the atom scattering length and
optical potential depth, are directly mapped to the Hubbard hopping
and on-site interaction terms. From the mapping diagram in
Ref.(\cite{georges}), it can be easily seen that most of the
parameter space investigated in this paper can be covered by the
present technology. Hence we are expecting a prompt experimental
realization of the transistor switch effect.

In summary, in this letter, we have put forward a novel scheme to
engineer the entanglement by using a three-site Hubbard model. The
whole idea is much like the traditional transistor switch, i.e. the
entanglement switching or tuning function can be turned on or off by
an external well-controllable parameter. Hence, we have not only
realized what QPT has been originally intended to do, but also gone
a step further by introducing an extra parameter to control its
functioning. The influence of the thermal fluctuations upon the
scheme has also been investigated. This scheme should have potential
applications in the QIP, especially for making entanglement control
devices, such as quantum memory or quantum storage media.

{\bf acknowledgements}

This work is supported by the National Basic Research Program of China (973 program) under Grant No.2006CB921104.\\

\newpage
\begin{figure}[htbp]
\centering
\includegraphics{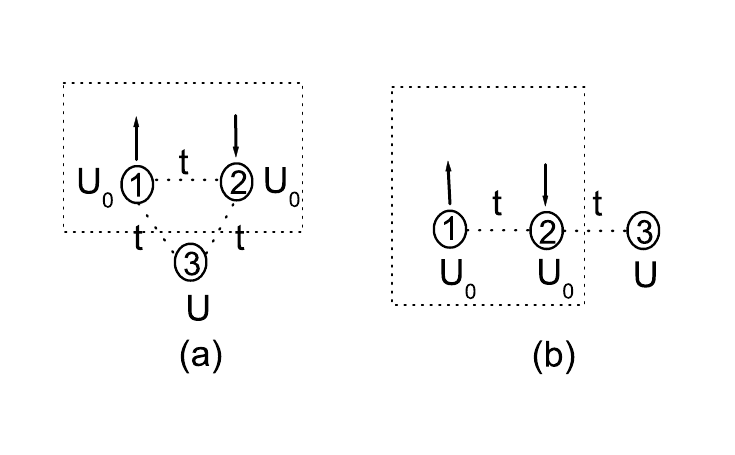}
\caption{Schematic diagram of the 3-site Hubbard model having two
electrons with different spins. (a) is for a cyclic case
corresponding to periodic boundary conditions and (b) for a linear
chain case corresponding to open boundary conditions. The
entanglement between the two sites in the dotted box are engineered
by varying $U/t$ and $U_0/t$. t is taken to be a scaling parameter
of the energy in our work.} \label{graph1}
\end{figure}

\begin{figure}
\centering
\includegraphics[width=\textwidth]{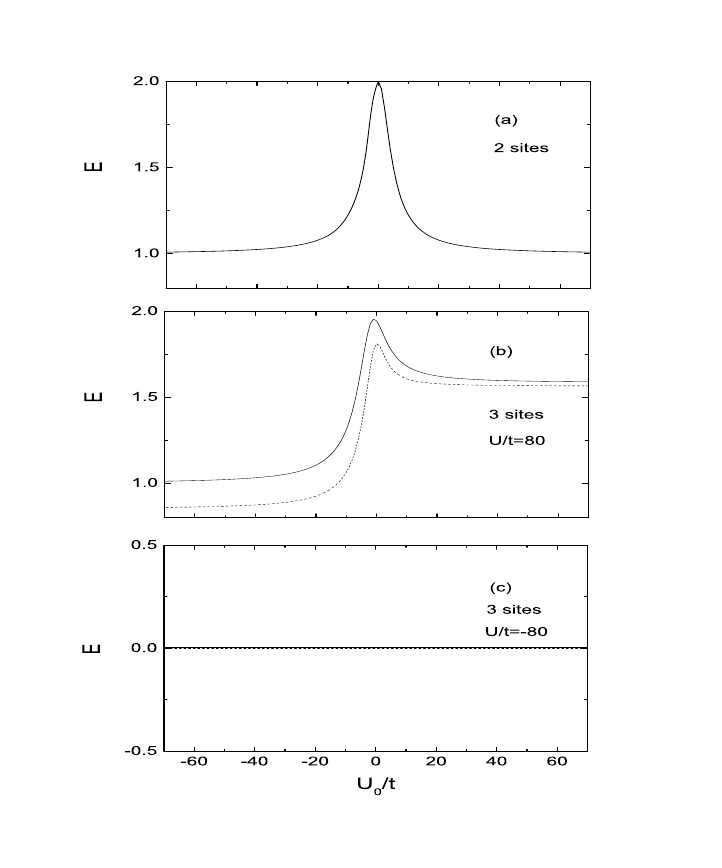}
\caption{Entanglement variations with respect to the tuning
parameter $U_0/t$. (a) is for the two-site case. (b) and (c) are for
the three-site case with positive and negative control parameter
$U/t$. In both (b) and (c), the solid line represents the cyclic
case and the dotted line the linear chain case.} \label{graph2}
\end{figure}

\begin{figure}[htbp]
\centering
\includegraphics[width=\textwidth]{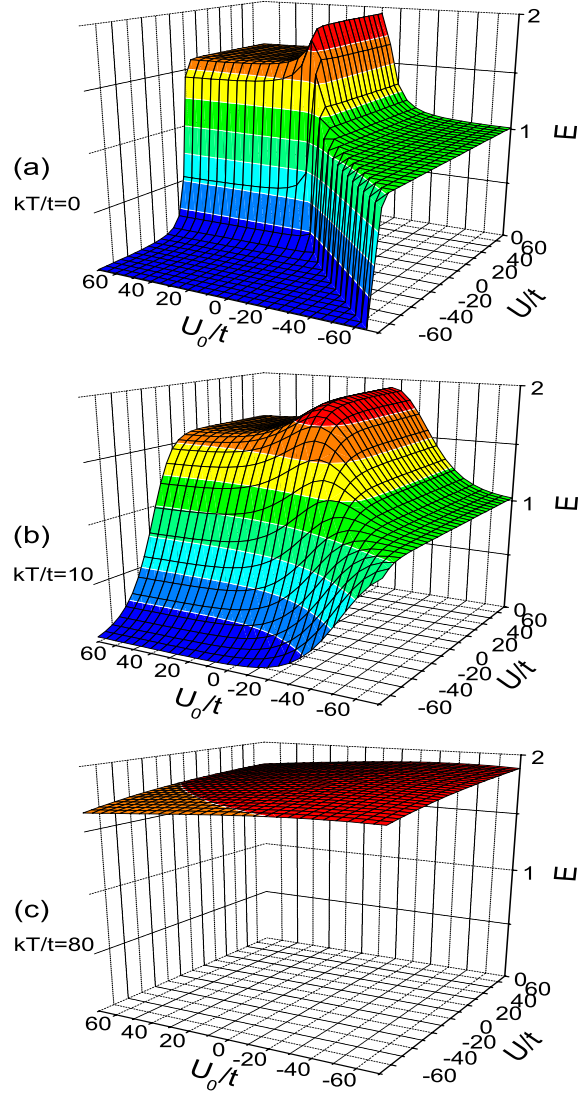}
\caption{Entanglement variations against $U/t$ and $U_0/t$ for the
3-site Hubbard model under different temperatures $kT$. (a)
$kT/t=0$. (b) $kT/t=10$. (c) $kT/t=80$.} \label{graph3}
\end{figure}

\end{document}